\documentstyle[pra,aps,twocolumn,psfig,floats,amssymb]{revtex}

\def\be{ \begin{equation} }
\def\ee{ \end{equation} }
\def\ba{ \begin{array} }
\def\ea{ \end{array} }
\def\bea{ \begin{eqnarray} }
\def\eea{ \end{eqnarray} }
\def\bml{ \begin{mathletters} }
\def\eml{ \end{mathletters} }
\def\bmla{ \bml \bea }
\def\emla{ \eea \eml }

\def\H{{\sf H}(t)}
\def\c{{\bf c}(t)}
\def\coupling{\alpha} 
\def\slope{\beta} 
\def\nl{\gamma} 
\def\D{{\cal D}}
\def\E{{\cal E}}

\def\J{{\cal J}}
\def\T{{\cal T}}
\def\a{\xi}
\def\NAC{\dot\vartheta} 

\begin{document}
\wideabs{
\draft
\author{N. V. Vitanov \protect\cite{email} and
	K.-A. Suominen \protect\cite{KASpermanent}}
\address{Helsinki Institute of Physics, PL 9,
 FIN-00014 Helsingin yliopisto, Finland}
\title{Nonlinear level crossing models}
\date{\today }
\maketitle

\begin{abstract}
We examine the effect of nonlinearity at a level crossing
on the probability for nonadiabatic transitions $P$.
By using the Dykhne-Davis-Pechukas formula, we derive simple analytic
estimates for $P$ for two types of nonlinear crossings.
In the first type, the nonlinearity in the detuning appears as a
{\it perturbative} correction to the dominant linear time dependence.
Then appreciable deviations from the Landau-Zener probability $P_{LZ}$
are found to appear for large couplings only, when $P$ is very small;
this explains why the Landau-Zener model is often seen
to provide more accurate results than expected.
In the second type of nonlinearity, called {\it essential} nonlinearity,
the detuning is proportional to an odd power of time.
Then the nonadiabatic probability $P$ is qualitatively and
quantitatively different from $P_{LZ}$ because on the one hand,
it vanishes in an oscillatory manner as the coupling increases,
and on the other, it is much larger than $P_{LZ}$.
We suggest an experimental situation when this deviation can be
observed.
\end{abstract}
}


\section{Introduction}
\label{Sec-introduction}

The Landau-Zener (LZ) model \cite{Landau32,Zener32} is widely used in quantum
physics to describe level crossing and avoided crossing transitions.
It provides the probability of transition between two quantum states coupled
by an external field of constant amplitude and time-dependent frequency
which passes through resonance with the transition frequency.
This {\it level crossing}, seen in the diabatic basis (the basis of the two
bare states), appears as an {\it avoided crossing} in the adiabatic basis
(the basis comprising the two eigenstates of the Hamiltonian).
Cases of level crossings and avoided crossings can be met in many
areas in physics, such as
laser-atom interactions \cite{Shore90},
magnetic resonance \cite{Abragam61},
slow \cite{Nikitin84} and cold \cite{Suominen96} atomic collisions,
molecular physics \cite{Child91,Garraway95},
optical atoms \cite{Spreeuw90,Bouwmeester95},
atom lasers \cite{Mewes97,Vitanov97},
solid-state physics \cite{Landauer85,Landauer87,Lenstra86,Gefen87},
ultrasmall tunnel junctions \cite{Likharev87,Mullen88,Schön90},
nuclear physics \cite{Imanishi87}, and
particle physics \cite{Haxton86,Parke86,Petcov87,Toshev87}.
The LZ model is a basic tool
for describing and understanding such phenomena.

Given that the LZ model presumes very crude time dependences
for the coupling (constant) and the detuning (linear),
it is somewhat surprising that it has been often found to provide rather
accurate results when applied to specific cases involving
time-dependent couplings (e.g., pulse-shaped) and nonlinear detunings.
To the best of our knowledge, no satisfactory explanation of this fact
has been given so far.
In the present paper, we try to answer this question by considering
models where the coupling is still constant but the detuning is
nonlinear.
We investigate two main classes of nonlinear detunings.
In the first class, the nonlinearity appears as a correction to a
dominant linear time dependence near the crossing; we call this
{\it perturbative} nonlinearity. 
In the second class, the detuning is proportional to an odd power of
time, $\Delta \propto t^N$, with $N=3,5,7,\ldots$; hence, the detuning
cannot be linearized in any vicinity of the crossing and we call this
nonlinearity {\it essential}.
We have found significant qualitative differences between these two cases.
We study the nonlinearity effects both numerically and analytically
and we are interested in the near-adiabatic regime which is the one of
primary interest as far as level crossing models are concerned.
The analytic approach is based on the Dykhne-Davis-Pechukas formula
\cite{Dykhne,Davis76} and a generalization of it that accounts
for multiple transition points
\cite{Davis76,George74,Suominen91,Suominen92pra,Suominen92oc}.

The paper is organized as follows.
The basic equations and definitions are given in Sec.~\ref{Sec-background}.
The perturbatively nonlinear models are discussed
in Sec.~\ref{Sec-perturbative}
and the essentially nonlinear models in Sec.~\ref{Sec-essential}.
The conclusions are summarized in Sec.~\ref{Sec-conclusion}.
The numerical method used for highly accurate integration of the
Schrödinger equation is nontrivial and it is presented in the Appendix.


\section{Background}
\label{Sec-background}

\subsection{Basic equations and definitions}

We wish to solve the Schr\"odinger equation ($\hbar=1$),
\be
\label{SEq}
i\frac d{dt} \c = \H \c,
\ee
with the Hamiltonian
\be
\label{H}
\H = \left[ 
\ba{cc}
-\Delta (t) & \Omega (t) \\ 
\Omega (t) & \Delta (t)
\ea
\right] ,
\ee
and $\c = \left[ c_1(t),c_2(t)\right]^T$, where $c_1(t)$ and $c_2(t)$
are the probability amplitudes of states $\psi _1$ and $\psi _2$.
The detuning $\Delta(t)$ and the coupling $\Omega(t)$ are both assumed real.
We are interested in the case when the coupling is constant
and the detuning is an odd function of time,
\be
\label{GeneralModel}
\Omega(t) = {\rm const}, \qquad \Delta(-t) = -\Delta(t).
\ee
Such is the case for the Landau-Zener (LZ) model, 
\be
\label{LZmodel}
\Omega(t) = {\rm const}, \qquad \Delta_{LZ}(t) = \slope^2 t.
\ee
Since the transition probability is invariant upon the sign inversions
$\Delta\rightarrow -\Delta$ and $\Omega \rightarrow -\Omega$, we
assume for simplicity and without loss of generality that both $\Omega$ and
the slope $\slope^2$ of the detuning at the crossing ($t=0$) are positive
and that $\slope $ itself is positive.

The level crossing, seen in the diabatic basis (\ref{SEq}), translates
into an avoided crossing in the adiabatic basis because the
eigenvalues $\pm \E(t)$ of the adiabatic states (the quasienergies)
do not cross but only come close to each other near $t=0$.
Here
\be
\label{E}
\E(t) = \sqrt{\Omega^2 + \Delta^2(t)}.
\ee

If the system is initially in state $\psi_1$
[$c_1(-\infty) = 1$, $c_2(-\infty) = 0$],
the transition probability in the diabatic basis at $+\infty$
is given by $\left| c_2(+\infty)\right|^2$.
The transition probability in the adiabatic basis (the probability for
nonadiabatic transitions) is equal to the probability of no transition
in the diabatic basis, $P = \left| c_1(+\infty)\right|^2$,
and its determination will be our main concern.
For the LZ model, 
\be
\label{Plz}
P=e^{-\pi \coupling ^2},
\ee
where $\coupling =\Omega /\slope $ is the normalized coupling.

\begin{figure}[tb]
\vspace*{0mm}
\centerline{\psfig{width=70mm,file=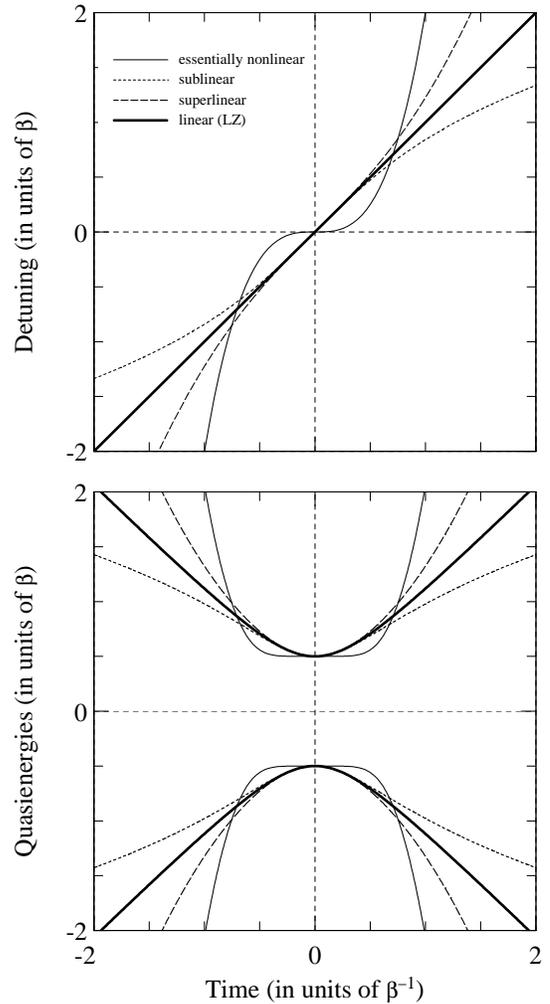}}
\vspace*{1mm}
\caption{
The level crossing cases studied in this paper.
Upper figure, the detunings $\Delta(t)$ in the diabatic basis.
Lower figure, the quasienergies $\pm\E(t)$ in the adiabatic basis.
}
\label{Fig-cases}
\end{figure}

In Secs.~\ref{Sec-perturbative} and \ref{Sec-essential}
we consider two types of nonlinear detunings and we
investigate how the nonlinearity affects the LZ probability (\ref{Plz}).
In the first type, the nonlinearity appears as a perturbative correction
to a dominant linear time dependence near the crossing;
for the superlinear model (Sec.~\ref{Sec-superlinear})
this correction is positive
whereas for the sublinear model (Sec.~\ref{Sec-sublinear})
the correction is negative.
In the second type, the detuning is proportional to $t^N$,
with $N=3,5,7,\ldots$, and the detuning is essentially nonlinear.
Like in the LZ model, the detuning diverges as $t\rightarrow\pm\infty$
in all considered cases.
The studied nonlinear detunings are shown in Fig.~\ref{Fig-cases}.

\subsection{Dykhne-Davis-Pechukas formula}

The Dykhne-Davis-Pechukas (DDP) formula \cite{Dykhne,Davis76}
provides the asymptotically exact probability for nonadiabatic
transitions in the adiabatic limit.
It reads
\be
\label{DP-1}
P \sim e^{-2{\rm Im}\D(t_c)},
\ee
where 
\be
\label{D(Tc)}
\D(t_c) = 2\int_0^{t_c}{\cal E}(t)dt.
\ee
The point $t_c$ is called the transition point and it is defined as the
(complex) zero of the quasienergy,%
\be
\label{Tc-def}
\E(t_c)=0, 
\ee
which lies in the upper complex $t$-plane (i.e., ${\rm Im}t_c>0$)
and if there are more than one such zero points,
$t_c$ is the one closest to the real axis.
Equation (\ref{DP-1}) gives the correct asymptotic probability for
nonadiabatic transitions provided
(i) the quasienergy $\E(t)$ does not vanish for real $t$,
including at $\pm \infty $;
(ii) $\E(t)$ is analytic and single-valued at least throughout a region
of the complex $t$-plane that includes the region
from the real axis to the transition point $t_c$;
(iii) the transition point $t_c$ is well separated from the other
quasienergy zero points (if any) and from possible singularities. 
Amazingly, for the LZ model, the DDP formula (\ref{DP-1}) gives the exact
probability (\ref{Plz}) not only in the adiabatic limit
($\coupling \gg 1$) but also for any $\coupling$.

In the case of more than one zero points in the upper $t$-plane,
Davis and Pechukas \cite{Davis76} have suggested,
following George and Lin \cite{George74},
that Eq.~(\ref{DP-1}) can be generalized to include the contributions
from all these $N$ zero points $t_c^k$ in a coherent sum.
This suggestion has been later verified by Suominen, Garraway and Stenholm
\cite{Suominen91,Suominen92pra,Suominen92oc,Suominen92}.
The generalized DDP formula has the form
\be
\label{DP-N}
P \sim \left| \sum_{k=1}^N \Gamma_k e^{i\D(t_c^k)}\right| ^2,
\ee
where
\be
\label{Gk}
\Gamma_k = 4i \lim \limits_{t\rightarrow t_c^k}(t-t_c^k) \NAC(t)
\ee
and $\NAC(t) \equiv [\dot\Omega(t)\Delta(t)-\Omega(t)\dot\Delta(t)]/2\E^2(t)$
is the nonadiabatic coupling, where an overdot means a time derivative.
In principle, Eq.~(\ref{DP-N}) should be used when
there are more than one zero points which are closest to the real axis
and have equal imaginary parts
and Eq.~(\ref{DP-N}) should include only the contributions from these zeroes.
The contributions from the farther zeroes are exponentially small
compared to the dominant ones and may therefore be neglected.
Retaining the contributions from {\it all} transition points, however,
may be beneficial and it has been shown elsewhere
\cite{Suominen92pra,Suominen92}
that for the Demkov-Kunike models \cite{Demkov69}
the full summation in Eq. (\ref{DP-N}) leads to the exact result.

The motivation for developing good approximations, such as
Eq.~(\ref{DP-N}) for the transition probability in the adiabatic
limit, does not arise only from the general wish to have analytic
expressions.
In fact, the numerical integration of the time-dependent Schrödinger
equation for the level crossing models studied in the present paper
(most of which are very slowly convergent from a numerical viewpoint)
becomes increasingly difficult and time consuming as we approach this limit.
In Appendix \ref{Sec-appendix}, we present the numerical approach used
in our studies, which has been specifically developed for highly
accurate numerical integration of slowly convergent two-state problems,
such as the present ones.
The need for sofisticated numerical approaches emphasizes the usefulness
of the DDP formulas (\ref{DP-1}) and (\ref{DP-N}), which become increasingly
accurate as the adiabatic limit is approached.


\section{Perturbative nonlinearity}
\label{Sec-perturbative}

We wish to estimate the probability for nonadiabatic transitions $P$ in the
case when the linear LZ detuning (\ref{LZmodel}) is perturbed by a small cubic
nonlinearity in the vicinity of the crossing (at $t=0$), 
\be
\label{perturbative}
\Omega(t) = {\rm const}, \qquad
\Delta(t) = \slope^2 t + \Gamma t^3 + \ldots.
\ee
When $\Gamma > 0$, the detuning passes through resonance
in a {\it superlinear} manner
($\left| \Delta (t)\right| >\left| \Delta _{LZ}(t)\right| $),
whereas when $\Gamma < 0$,
the detuning passes through resonance in a {\it sublinear} manner
($\left| \Delta (t)\right| <\left| \Delta _{LZ}(t)\right| $).
It is more convenient to work with dimensionless quantities
and we choose $\slope$ to define our frequency and time scales.
We define the dimensionless coupling $\coupling$, the nonlinearity
coefficient $\nl$, and the time $\tau$ as
\be
\label{parameters}
\coupling = \frac \Omega \slope, \qquad
\nl = \frac {\left| \Gamma \right|^{\frac12}} {\slope^2}, \qquad
\tau = \slope t,
\ee
both $\coupling$ and $\nl$ being positive.
Then Eqs. (\ref{perturbative}) become
\be
\label{perturbative-dimless}
\widetilde{\Omega}(\tau) = \coupling, \qquad
\widetilde{\Delta}(\tau)
	= \tau \left(1 \pm \nl^2 \tau^2 + \ldots \right),
\ee
where $\widetilde{\Omega}(\tau) = \Omega(t) /\slope$,
$\widetilde{\Delta}(\tau) = \Delta(t) /\slope$,
and the plus (minus) sign is for the superlinear (sublinear) case.

The direct treatment of model (\ref{perturbative-dimless}) involves
a third-order algebraic equation which is too cumbersome.
Moreover, for a negative nonlinearity, model (\ref{perturbative-dimless})
involves two unwanted additional spurious crossings.
We avoid these drawbacks of model (\ref{perturbative-dimless})
by using instead two other models which contain positive and negative
cubic nonlinearities, have a single level crossing, involve dealing
with quadratic equations, and allow simpler derivations.


\subsection{Superlinear model}
\label{Sec-superlinear}

The first model we consider is 
\be
\label{superlinear}
\widetilde{\Omega}(\tau) = \coupling, \qquad
\widetilde{\Delta}(\tau) = \tau \sqrt{1 + 2\nl^2 \tau^2}.
\ee
For $\nl=0$, model (\ref{superlinear}) reduces to the LZ model.
Near the crossing ($\tau=0$), we have
$\widetilde{\Delta}(\tau) = \tau \left(1 + \nl^2 \tau^2 + \ldots \right)$,
i.e., model (\ref{superlinear}) reduces
to model (\ref{perturbative-dimless}) with a {\it positive} nonlinear term. 
Moreover, as $\tau \rightarrow \pm \infty $, the detuning diverges
quadratically, i.e., {\it faster} than the (linear) LZ detuning.
Hence, we call model (\ref{superlinear}) the {\it superlinear model}.

This model has two transition points in the upper complex half-plane,
\be
\label{Tc-super}
\tau_c^\pm = \frac {i\coupling}{\xi} \left(\sqrt{1+\a} \pm \sqrt{1-\a}\right),
\ee
where $\a \equiv 2\sqrt{2}\coupling \nl$.
For $\nl \rightarrow 0$, we have $\tau_c^- \rightarrow i\coupling$
and $\tau_c^+ \rightarrow i\infty$,
and we recover the (single) LZ transition point.

\subsubsection{The case of small $\coupling \nl$}

For $\a \equiv 2\sqrt{2}\coupling \nl < 1$, both transition points
are purely imaginary, with ${\rm Im}\tau_c^- < {\rm Im}\tau_c^+$.
Hence, we can take the contribution from $\tau_c^-$ only.
We make the substitutions $\tau = iy$ and $\tau_c^- = iy_c$
in Eq.~(\ref{D(Tc)}) and obtain
\bea
\label{D-super-small}
&& \D\left(\tau_c^-\right) = 2i \int_0^{y_c}
	\sqrt{\coupling^2 - y^2 + 2\nl^2 y^4}\ dy \nonumber\\
&& \qquad = \case12 \pi\coupling\tau_c^-
F\left(\case12,-\case12;2;\frac {1-\sqrt{1-\xi^2}}{1+\sqrt{1-\xi^2}}\right)
	\nonumber\\
&& \qquad = i \frac{\pi\coupling^2}{2} F(\case14,\case34;2;\xi^2),
\eea
where $F(a,b;c;z)$ is the Gauss hypergeometric function \cite{AS}.
The DDP formula gives
\be
\label{Psuper-small}
P\sim e^{-\pi\coupling^2 F(\frac14,\frac34;2;\xi^2)}. 
\ee
For $\a \equiv 2\sqrt{2}\coupling \nl \ll 1$, Eq.~(\ref{Psuper-small})
reduces to
\be
\label{Psuper-approx}
P \sim e^{-\pi \coupling ^2\left(1+\frac34\coupling^2\nl^2+\cdots\right) }.
\ee
Comparison with the LZ formula (\ref{Plz}) shows that the superlinearity
{\it reduces} the probability for nonadiabatic transitions.

The DDP result (\ref{Psuper-approx}) is formally valid in the near-adiabatic
regime ($\coupling \gg 1$).
Moreover, in the derivation of Eq.~(\ref{Psuper-approx}) we
assumed that $2\sqrt{2}\coupling \nl \ll 1$.
Hence, we should have
\be
\label{super-condition}
\frac 1 {2\sqrt{2}\nl} \gg \coupling \gg 1.
\ee
We have to account also for the fact that both the other zero point $\tau_c^+$
and the singularity of model (\ref{superlinear}), situated at
$\tau_0 = i/\nl\sqrt{2}$,
should be well separated from the transition point $\tau_c^-$.
It is easy to verify that in the present case
($2\sqrt{2}\coupling \nl < 1$) we have
${\rm Im}\tau_c^- < {\rm Im}\tau_c^+ < {\rm Im}\tau_0$ and hence,
we should only have ${\rm Im}\tau_c^+ - {\rm Im}\tau_c^- \gg 1$.
This is indeed the case if condition (\ref{super-condition}) is satisfied.

Condition (\ref{super-condition}) limits the applicability
of Eq.~(\ref{Psuper-approx}) to very small nonlinearities only.
We have found, however, through extensive comparison with numerical
simulations that this condition can be relaxed to
\be
\frac 1{2\sqrt{2}\nl} \gtrsim \coupling \gtrsim 1.
\ee

\subsubsection{The case of large $\coupling \nl$}

For $\a \equiv 2\sqrt{2}\coupling\nl >1$, the transition points
(\ref{Tc-super}) are complex, with equal imaginary parts
and real parts which are equal in magnitude but opposite in sign.
Then, following Eq.~(\ref{DP-N}),
we have to take the contributions from both of them.
The integrals $\D(\tau_c^\pm)$ are given by
\bea
&& \D(\tau_c^\pm) = \case12 \pi\coupling\tau_c^\pm
F\left(\case12,-\case12;2;\frac {1\pm i\sqrt{\xi^2-1}}{1\mp i\sqrt{\xi^2-1}}
	\right)\nonumber\\
&& \qquad = \pm\D_r+i\D_i,
\eea
where $\pm\D_r$ and $\D_i$ are the real and imaginary parts.
The factors (\ref{Gk}) are $\Gamma_\pm=-1$ and
the generalized DDP formula (\ref{DP-N}) gives
\be
\label{Psuper-large}
P \sim 4 e^{-2\D_i} \cos^2 \D_r . 
\ee

\begin{figure}[tb]
\vspace*{0mm}
\centerline{\psfig{width=75mm,file=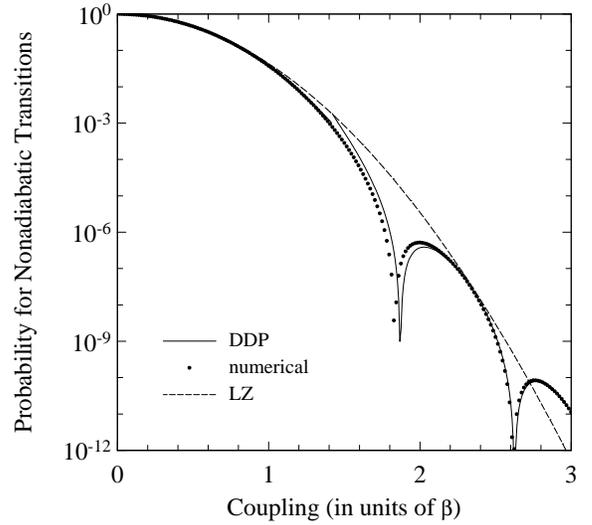}}
\vspace*{2mm}
\caption{
The probability for nonadiabatic transitions $P$
for the {\it superlinear} model (\protect\ref{superlinear})
plotted against the dimensionless coupling $\coupling \equiv \Omega/\slope$
for $\nl=0.25$.
The numerical results are shown by dots
while the DDP approximations (\protect\ref{Psuper-approx})
(for $\coupling\leqq\protect\sqrt{2}$) and (\protect\ref{Psuper-large})
(for $\coupling>\protect\sqrt{2}$) by solid curves
[the small-$\coupling$ approximation (\protect\ref{Psuper-approx})
is almost indistinguishable from the numerical results].
The LZ probability (\protect\ref{Plz}) is shown by a dashed curve.
}
\label{Fig-superlinear}
\end{figure}

In Fig.~\ref{Fig-superlinear}, the probability for nonadiabatic transitions
for the superlinear model (\protect\ref{superlinear}) is plotted against
the dimensionless coupling $\coupling$ for nonlinearity $\nl=0.25$.
The DDP approximations (\protect\ref{Psuper-approx})
and (\protect\ref{Psuper-large}) are seen to fit the numerical results
very well.
As seen from the figure, this is the case even for small $\coupling$,
while the DDP formula is supposed to be valid for $\coupling \gtrsim 1$ only.
This is a consequence of the fact that the DDP formula provides
the exact result for the LZ model for any $\coupling$.

The condition of validity of Eq.~(\ref{Psuper-large}) is expected to be
$\a=2\sqrt{2}\coupling\nl \gtrsim 1$.
However, for large coupling $\coupling$ and nonlinearity $\nl$,
the singularity point $\tau_0 = i/\nl\sqrt{2}$ gets closer to the real
axis than the transition points $\tau_c^\pm$ which makes the DDP
result (\ref{Psuper-large}) inaccurate for very large $\coupling\nl$;
indeed, we have verified this numerically
(not shown in Fig. \ref{Fig-superlinear}).
We should emphasize that the case of large nonlinearity resembles more that
of the essential nonlinearity (Sec.~\ref{Sec-essential}) and hence, it is
not interesting in the context of {\it perturbative} nonlinearity,
considered in this section.

\subsection{Sublinear model}
\label{Sec-sublinear}

The sublinear model is defined as
\be
\label{sublinear}
\widetilde{\Omega}(\tau) = \coupling, \qquad
\widetilde{\Delta}(\tau) = \frac{\tau}{\sqrt[4]{1+4\nl^2 \tau^2}}. 
\ee
For $\nl =0$, it reduces to the LZ model (\ref{LZmodel}).
Near the crossing the detuning behaves as
$\widetilde{\Delta}(\tau) = \tau \left(1-\nl^2\tau^2+\ldots\right)$,
i.e., model (\ref{sublinear}) reduces to model (\ref{perturbative-dimless})
with a negative nonlinear term.
As $\tau \rightarrow \pm \infty$, the detuning diverges as $|\tau|^{\frac12}$,
i.e., {\it slower} than the (linear) LZ detuning.
Hence, the name {\it sublinear} model.

This model has a single transition point in the upper half plane given by
\be
\label{Tc-sub}
\tau_c = i \coupling \sqrt{\sqrt{4\coupling^4\nl^4+1}-2\coupling^2\nl^2}. 
\ee
The integral $\D\left(\tau_c\right)$ is given by
\be
\D\left( \tau_c\right) = i{\J}, 
\ee
where
\be
\J = 2 \int_0^{y_c} \sqrt{\coupling^2 - \frac{y^2}{\sqrt{1-4\nl^2y^2}}}\ dy. 
\ee
where the substitutions $\tau = iy$ and $\tau_c = i y_c$ have been made.
The DDP formula (\ref{DP-1}) gives
\be
\label{Psublinear}
P \sim e^{-2\J}.
\ee
In the limits of small and large $\coupling\nl$, we find
\bml
\label{Psub-approx}
\bea
\label{Psub-small}
&& P \sim e^{-\pi \coupling^2
	\left( 1-\frac34\coupling^2\nl^2+\cdots\right) },\ \ 
	(\sqrt{2}\coupling \nl \lesssim 1), \\
\label{Psub-large}
&& P \sim e^{-2\coupling/\nl + \pi/(16\coupling\nl^3)+\cdots},\ \ 
	(\sqrt{2}\coupling \nl \gtrsim 1).
\eea
\eml
Comparison with the LZ formula (\ref{Plz}) shows that the sublinearity
{\it increases} the probability for nonadiabatic transitions.
Note also that due to the absence of other transition points in the upper
half-plane, there are no oscillations in $P$.

\begin{figure}[tb]
\vspace*{0mm}
\centerline{\psfig{width=75mm,file=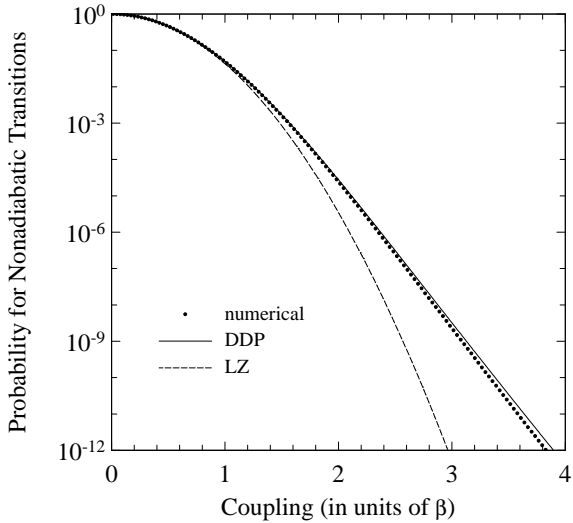}}
\vspace*{2mm}
\caption{
The probability for nonadiabatic transitions $P$ for the {\it sublinear} model
(\protect\ref{sublinear}) plotted against the dimensionless coupling
$\coupling \equiv \Omega/\slope$ for $\nl=0.25$.
The numerical results are shown by dots and
the analytic approximation (\protect\ref{Psublinear}) by a solid curve.
The LZ probability (\protect\ref{Plz}) is shown by a dashed curve.
}
\label{Fig-sublinear}
\end{figure}

In Fig.~\ref{Fig-sublinear}, the probability for nonadiabatic transitions
$P$ for the sublinear model (\protect\ref{sublinear}) is plotted against
the dimensionless coupling $\coupling$ for nonlinearity $\nl=0.25$.
The DDP approximation (\protect\ref{Psublinear}) is seen to fit
the numerical results very well.

The condition of validity of the DDP approximation (\ref{Psublinear})
is expected to be $\coupling \gtrsim 1$.
In fact, as Fig.~\ref{Fig-sublinear} shows, Eq. (\ref{Psublinear}) is
quite accurate for $\coupling<1$ too, which is related to the fact the
DDP formula provides the exact LZ probability, as noted above.
On the other hand, we have to account for the presence of a singularity
at $\tau_0 = i/2\nl$.
The relation ${\rm Im}\tau_c < {\rm Im}\tau_0$ is always fulfilled
which means that the DDP formula (\ref{Psublinear})
should be accurate if the two points $\tau_c$ and $\tau_0$ are well separated.
For large $\coupling\nl$, however, they approach each other
[see Eq.~(\ref{Tc-sub})],
which explains the small inaccuracy of Eq.~(\ref{Psublinear}) seen in
Fig.~\ref{Fig-sublinear}.


\subsection{Discussion}

Comparison of Eqs. (\ref{perturbative-dimless}), (\ref{Psuper-approx}) and
(\ref{Psub-small}) shows that the probability for nonadiabatic transitions
for model (\ref{perturbative}) for small nonlinearity is given by
\be
\label{Pperturbative}
P \sim P_{LZ} e^{-\frac34 \pi \Omega^4 \Gamma / \slope^8 }.
\ee
This equation is valid for
$\Omega^2\Gamma/\slope^6 \lesssim 1$.
Since the deviation from the LZ probability $P_{LZ}$
depends exponentially on the coupling $\Omega$
and the nonlinearity coefficient $\Gamma$,
this deviation can be very large.
On the other hand, however, this difference emerges when
$\Omega/\slope \gtrsim 1$; then $P$ is very small (virtually zero),
which explains why the LZ formula is often found
to be more accurate than anticipated.

\begin{figure}[tb]
\vspace*{0mm}
\centerline{\psfig{width=75mm,file=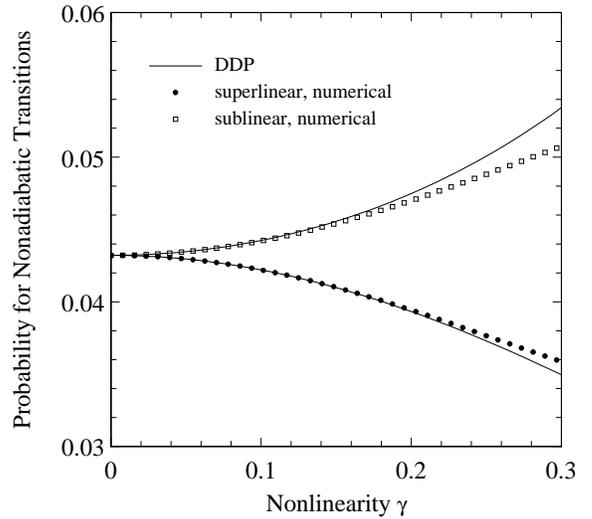}}
\vspace*{2mm}
\caption{
The probability for nonadiabatic transitions $P$
plotted against the dimensionless nonlinearity coefficient $\nl$
for the superlinear model (\protect\ref{superlinear}) (full circles)
and the sublinear model (\protect\ref{sublinear}) (open squares).
The coupling is $\coupling \equiv \Omega/\slope = 1$.
The solid curves depict the analytic approximation
(\protect\ref{Pperturbative}) derived from the DDP formula.
}
\label{Fig-nonlinearity}
\end{figure}

In Fig.~\ref{Fig-nonlinearity},
the probability for nonadiabatic transitions $P$
is plotted against the nonlinearity coefficient $\nl$
for the superlinear model (\ref{superlinear})
and the sublinear model (\ref{sublinear}).
For $\nl=0$, both the superlinear and sublinear probabilities are equal
to the LZ value $P_{LZ} \approx 0.043$.
As $\nl$ departs from zero, the superlinear probability decreases while
the sublinear probability increases, in agreement with our analytic
result (\ref{Pperturbative}).
This qualitatively different behavior is readily explained by looking at
the quasienergies $\pm\E(t)$.
For the superlinear model (\ref{superlinear}), the avoided crossing
between the quasienergies is {\it sharper} than in the LZ case, while
for the sublinear model (\ref{sublinear}), it is {\it flatter}
than the LZ one (see Fig.~\ref{Fig-cases}).
Consequently, the ``duration'' $\T$ of the avoided crossing for the
superlinear model is shorter than for the LZ model, while for the sublinear
model it is larger, $\T_{super}<\T_{LZ}<\T_{sub}$, which means that the same
relation should apply for the probabilities, $P_{super}<P_{LZ}<P_{sub}$.

\subsection{Comparison with the Allen-Eberly-Hioe model}

An argument in favor of Eq.~(\ref{Pperturbative}) can
be derived from the Allen-Eberly-Hioe model \cite{Allen75,Hioe84},
\be
\label{AEHmodel}
\Omega(t) = \frac A {\cosh(t/T)}, \qquad
\Delta(t) = B \tanh(t/T),
\ee
which is a particular case of the Demkov-Kunike model
\cite{Suominen92pra,Demkov69}.
Model (\ref{AEHmodel}) can be solved analytically and gives
\be
\label{Paeh}
P = \frac {\cosh^2\left(\pi\sqrt{B^2-A^2} T \right)} {\cosh^2(\pi BT)}. 
\ee
It has been shown in \cite{Vitanov96} that this model is a member
of a class of infinite number of models, for all of which
the nonadiabatic probability is given by Eq.~(\ref{Paeh}).
Another member of this class is the model
\bea
\label{AEHmodel-1}
&&\Omega(t) = A,\qquad \Delta(t) = B \tan(t/T), \\
&&\qquad (-\case12 \pi T\leqq t \leqq \case12 \pi T).\nonumber
\eea
We can make model (\ref{AEHmodel-1}) behave like the perturbatively
nonlinear model (\ref{perturbative})
(i.e., make the coupling duration infinite and the detuning divergent)
by letting $T\rightarrow\infty$ and $B\rightarrow\infty$,
while maintaing the slope at the crossing $B/T$ constant.
Then Eq.~(\ref{Paeh}) reduces to
\be
\label{Paeh-approx}
P \sim \exp\left[-\pi\frac{A^2 T}{B}
	\left(1 + \frac {A^2}{4B^2} + \ldots \right)\right]. 
\ee
The same result follows from the DDP approximation (\ref{Pperturbative})
by accounting for the fact that for model (\ref{AEHmodel-1}),
$\coupling=A\sqrt{T/B}$, $\slope^2=B/T$, and $\Gamma=B/3T^3$.


\section{Essential nonlinearity}
\label{Sec-essential}

The essentially nonlinear model is defined by
\be
\label{essential}
\Omega(t) ={\rm const}, \qquad \Delta(t) = \slope^{N+1} t^N, 
\ee
where $N=3,5,7,\ldots$ is an odd number.
Model (\ref{essential}) cannot be linearized in any vicinity of the crossing.

The zero points of the quasienergies (\ref{E}) in the upper $t$-plane are
\be
\label{Tc-essential}
\tau_c^k = \coupling^{1/N} e^{i\pi (2k-1)/(2N)},
\qquad (k=1,2,\ldots ,N), 
\ee
where $\tau=\slope t$ and $\coupling=\Omega/\slope$.
With the exception of $\tau_c^{(N+1)/2}$,
which is imaginary, the transition points are grouped in pairs that have the
same imaginary parts but opposite real parts.
The significant difference between the LZ model and the essentially nonlinear
model is that $\tau_c^{(N+1)/2}$ is the only transition point in the former,
while in the latter it is the one farthest from the real axis and hence,
with the smallest contribution to the nonadiabatic probability.
The largest contribution comes from the points $\tau_c^1$ and $\tau_c^N$
which are closest to the real axis.

The integrals $\D(\tau_c^k)$ are given by
\be
\D(\tau_c^k) = 2\nu_N \coupling ^{(N+1)/N}e^{i\pi (2k-1)/(2N)}, 
\ee
where the number $\nu_N$ denotes the value of the integral
\be
\nu_N = \int_0^1 \sqrt{1-x^{2N}}\ dx.
\ee
The first few values of $\nu_N$ are
$\nu_1 = \pi /4\approx 0.785$,
$\nu_3 \approx 0.911$,
$\nu_5 \approx 0.944$,
$\nu_7 \approx 0.959$.
The factors (\ref{Gk}) are given by $\Gamma_k = (-1)^k$.
The generalized DDP formula (\ref{DP-N}) gives 
\bea
\label{Pessential}
&& P \approx \bigg[2 \sum_{k=1}^{(N-1)/2} (-1)^k
	e^{-\eta \sin \frac{2k-1}{2N}\pi}
	\cos\left(\eta \cos \frac{2k-1}{2N}\pi\right)\nonumber\\
&&\qquad + (-1)^{(N+1)/2} e^{-\eta} \bigg]^2,
\eea
with $\eta=2\nu_N \coupling^{(N+1)/N}$.
For large $\coupling$, the main contribution to $P$ comes from the first
term ($k=1$) in the sum.

\begin{figure}[tb]
\vspace*{0mm}
\centerline{\psfig{width=75mm,file=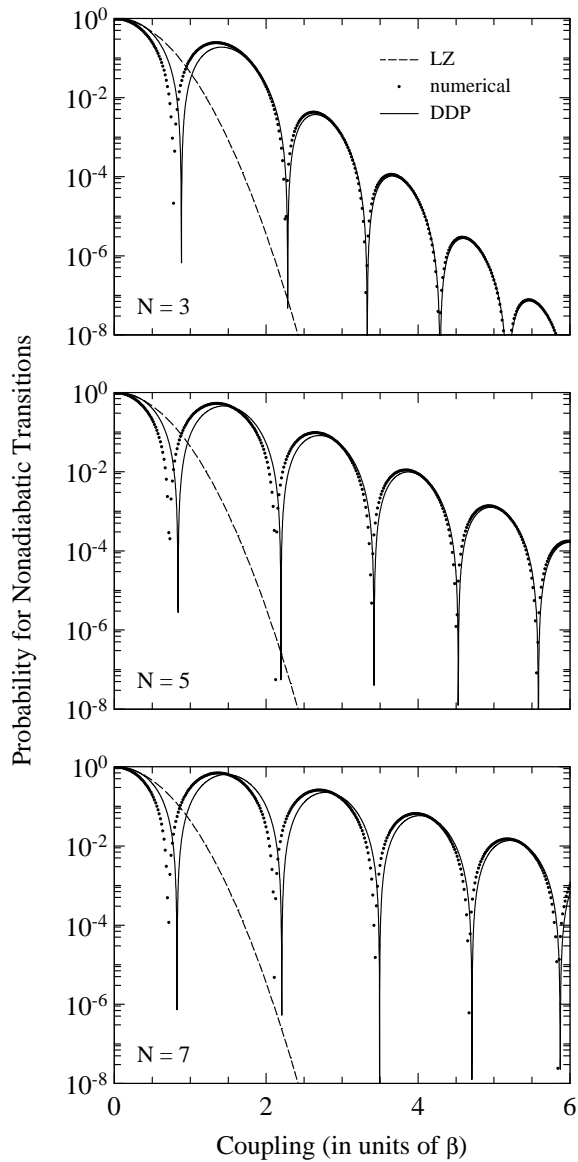}}
\vspace*{2mm}
\caption{
The probability for nonadiabatic transitions $P$
for the essentially nonlinear crossing model (\protect\ref{essential})
plotted against the dimensionless coupling $\coupling \equiv \Omega/\slope$
for $N=3$ (top figure), $N=5$ (middle figure), and $N=7$ (bottom figure).
In each figure, the numerical results are shown by dots,
the analytic approximation (\protect\ref{Pessential}) by a solid curve,
and the LZ probability (\protect\ref{Plz}) by a dashed curve.
}
\label{Fig-essential}
\end{figure}

The probability (\ref{Pessential}) is
plotted in Fig.~\ref{Fig-essential} against the coupling $\coupling$
and compared with the exact numerical results
for $N=3$ (top figure), $N=5$ (middle figure), and $N=7$ (bottom figure).
Obviously, the probability (\ref{Pessential})
is qualitatively different from the LZ probability $P_{LZ}$.
First, there are oscillations which appear due to the existence of
multiple transition points.
From another point of view, the oscillations appear because the
nonadiabatic coupling $\dot\vartheta(t)$ has two peaks, in contrast to
the single-peaked nonadiabatic coupling in the LZ model.
Second, the transition probability is much larger than $P_{LZ}$
which can be explained by the fact that in the quasienergy picture,
the avoided crossing has a much longer duration (see Fig.~\ref{Fig-cases}).

These differences can be verified experimentally.
For example, an essentially nonlinear crossing arises when a two-state
atom is excited by a frequency modulated laser pulse with a supergaussian
time dependence [i.e., $\Omega(t) \propto e^{-(t/T)^{2n}}$] and the
frequency modulation is produced by the self-phase modulation technique,
in which the phase shift $\varphi(t)$ is proportional to either the amplitude
of the field [i.e., $\varphi(t)\propto \Omega(t)$] or the intensity
of the field [i.e., $\varphi(t)\propto \Omega^2(t)$] \cite{Goswami94}.
Since the detuning is proportional to $\dot\varphi(t)$,
we shall have $\Delta(t)\propto t^{2n-1}$ with a crossing at $t=0$.
As the hypergaussians are almost constant near $t=0$,
this example reproduces model (\ref{essential}) very well.


\section{Summary and conclusions}
\label{Sec-conclusion}

We have examined the effect of nonlinearity in the detuning at the level
crossing on the probability for nonadiabatic transitions $P$.
Our analysis has been based upon both analytic approximations derived
by using the Dykhne-Davis-Pechukas formula and  numerical calculations.
We have distinguished two types of nonlinearities: perturbative and
essential.
In the former type, the nonlinearity appears as a correction to a
dominant linear time dependence near the crossing. 
In the latter, the detuning is proportional to an odd power of
time, $\Delta \propto t^N$ with $N=3,5,7,\ldots$.

For the {\it perturbative nonlinearities}, the probability for nonadiabatic
transitions $P$ is larger than $P_{LZ}$ for sublinear nonlinearity and
smaller than $P_{LZ}$ for superlinear nonlinearity.
For both the superlinear and sublinear cases, the appreciable deviations
from $P_{LZ}$ emerge for large coupling
when $P$ is already very small, virtually unobservable.
This fact explains why the LZ model has often been found to provide more
accurate results than anticipated.
We have provided a simple analytic estimate for the deviation as a
function of the nonlinearity, which should be a useful criterion for
estimating the applicability of the LZ model to any level crossing case,
as far as perturbative nonlinearity is concerned.

For the {\it essential nonlinearity}, we have found that
the nonadiabatic probability $P$ is both quantitatively and qualitatively
different from $P_{LZ}$ because on the one
hand, it vanishes in an oscillatory manner as the coupling increases,
and on the other, it is much larger than $P_{LZ}$.
From a mathematical viewpoint, these differences derive from the
existence of two complex transition points equally close to the real
axis, both being closer than the single LZ transition point.
From a physical viewpoint, the differences can be explained by the fact
that the avoided crossing is flatter and of longer duration than the LZ one
and the nonadiabatic coupling has two peaks, in contrast to
the single-peaked LZ nonadiabatic coupling.
We have suggested an experimental situation when this deviation can be
observed.

We have limited our analysis to the case when the detuning is an odd
function of time [Eq.~(\ref{GeneralModel})], which excludes asymmetric
level crossings, e.g., quadratic corrections in the perturbative models.
It would be interesting to estimate the effect of such quadratic terms
although then the DDP treatment is more complicated.
It should be noted that the parabolic level crossing model considered in
\cite{Suominen92oc} involves either two level crossings or no crossing
and hence, it is different from the present case of a single level crossing.

Finally, we point out that the detuning nonlinearity is only one
of the sources of possible inaccuracies in applications of the LZ model.
Others include the finite coupling duration \cite{Vitanov96} and
the nonzero transition times \cite{Vitanov99}.


\subsection*{Acknowledgments}

This work has been supported financially by the Academy of Finland.


\begin{appendix}

\section{Numerical integration of slowly convergent two-state problems}
\label{Sec-appendix}

In this Appendix, we describe the numerical algorithm we have used to
integrate the Schrödinger equation~(\ref{SEq}).
The Landau-Zener model is notoriously known for its slow numerical
convergence because the amplitude of the oscillations, which appear in the
time evolution of the transition probability $P$, only vanishes
as $|t|^{-1}$ in the diabatic basis \cite{Vitanov96}.
This is a consequence of the slow (linear) divergence of the detuning.
Of course, the final LZ transition probability (\ref{Plz}) is known exactly,
but numerical integration is necessary when the time evolution is needed.
The situation is even worse for the sublinear model (\ref{sublinear})
in which the oscillations amplitude vanishes only as $|t|^{-\frac12}$.
One of the possibilities to alleviate this problem is to carry out the
numerical integration in the adiabatic basis, where the oscillation
amplitude vanishes as $|t|^{-3}$ for the LZ model \cite{Vitanov99} and as
$|t|^{-2}$ for the sublinear model.
This improvement may be insufficient if the transition probability
is very small, as in the present paper.
Yet another problem arises from the finite initial (large negative)
time, because it introduces additional oscillations in $P$ \cite{Vitanov96}.
For the LZ model these problems can be resolved by starting the integration
at the crossing ($t=0$) and propagating the solution towards the desired
(positive or negative) time \cite{Vitanov96,Vitanov99}.
The initial conditions in this approach require the values of $P$ and
a few of its derivatives at $t=0$, which can be found exactly
\cite{Vitanov96,Vitanov99}.
Unfortunately, this approach does not apply to the models in the present paper.

We have used a numerical approach which combines and generalizes ideas
by Bambini and Lindberg \cite{Bambini84} and Berry and Lim \cite{Berry93}.
The approach is based on two concepts.
First, we perform the numerical integration in a superadiabatic (SA) basis
\cite{Berry93}, rather than in the usual diabatic or adiabatic bases.
The successive SA bases are obtained iteratively \cite{Berry93,Berry87,Drese98}
and they are not the more familiar superadiabatic bases obtained as
truncated asymptotic series in the adiabatic parameter \cite{Berry90}.
The $n$-th order SA states are defined as the instantaneous normalized
eigenstates of the Hamiltonian in the SA basis of order $n-1$.
For instance, by diagonalizing the Hamiltonian in the diabatic basis
(which is the SA basis of order $n=0$) we obtain the adiabatic basis (SA
basis of order $n=1$).
The recursive relations between the ``couplings'' $\Omega_n(t)$ and the
``detunings'' $\Delta_n(t)$ in two successive SA bases are given by
\bmla
&& \Delta_{n+1}(t) = \sqrt{\Omega_n^2(t) + \Delta_n^2(t)}, \\
&& \Omega_{n+1}(t) = \frac{\dot{\Omega}_n(t)\Delta_n(t)
			   -\Omega_n(t)\dot{\Delta}_n(t)}
		      {2\Delta_{n+1}^2(t)}.
\emla
Here $\Delta_0(t)\equiv \Delta(t)$ and $\Omega_0(t)\equiv \Omega(t)$
are the detuning and the coupling in the diabatic basis, whereas
$\Delta_1(t) \equiv \E(t)$ and $\Omega_1(t) \equiv \dot{\vartheta}(t)$ are the
quasienergy and the nonadiabatic coupling in the adiabatic basis.
As far as the level crossing models in the present paper are concerned,
the advantage of using the $n$-th SA basis is that the oscillations in
the transition probability evolution, whose amplitude is proportional to
the ratio $\Omega_n(t)/\Delta_n(t)$ at large times, vanish much more quickly.
This is so because at large times, the SA ``detuning'' $\Delta_n(t)$
diverges in the same manner as $\Delta(t)$, while
the SA ``coupling'' $\Omega_n(t)$ vanishes as $|t\Delta(t)|^{-n}$.
Hence, the oscillation amplitude vanishes as $|t|^{-n}|\Delta(t)|^{-n-1}$.
We have used the third SA basis ($n=3$), in which the oscillation amplitude
vanishes as $|t|^{-7}$ for the LZ model (\ref{LZmodel}),
as $|t|^{-5}$ for the sublinear model (\ref{sublinear}),
as $|t|^{-11}$ for the superlinear model (\ref{superlinear}),
and as $|t|^{-4N-3}$ for the essentially nonlinear models (\ref{essential}).
Moreover, this approach provides the possibility to check the accuracy
by calculating $P$ in different SA bases.

As we have pointed out above, another problem, which cannot be resolved
merely by the choice of basis, is the finite initial time.
We have overcome it in the manner of Bambini and Lindberg
\cite{Bambini84} by using the symmetry of the two-state
problem in the $n$-th SA basis.
The Bambini-Lindberg approach is based on a connection between
the two-state evolution matrix ${\bf U}(t,0)$,
describing the evolution from time $t=0$ to time $t$,
and the evolution matrix ${\bf U}(t,-t)$,
describing the evolution from $-t$ to $t$.
This approach allows to start the integration at $t=0$, propagate it
towards $+\infty$, and stop the integration when some convergence
criterion is fulfilled [for instance, we have required that the
oscillation amplitude at time $t$ is smaller than $10^{-4}P(t)$].
It is easy to see that this approach is much faster than merely twice
compared to the standard one (starting at large negative time), in which a
convergence check would require to start the entire integration again at a
larger negative time.
The Bambini-Lindberg approach needs to be generalized because
it applies to the case of a coupling and a detuning that are both
even functions of time, while in the present case,
the $n$-th SA ``detuning'' is an odd function in the diabatic basis
and an even function in the SA bases, while the $n$-th SA ``coupling''
can be either an even function (for odd $n$)
or an odd function (for even $n$).
The relations between ${\bf U}(t,-t)$ and ${\bf U}(t,0)$ for the four
possible combinations of symmetries in the detuning and the coupling
(odd or even functions) have been derived in \cite{Vitanov94}
in a similar manner as in \cite{Bambini84}.
We present the results in Table \ref{Table}.
In the table, $b_1 \equiv b_1(+\infty)$ and $b_2 \equiv b_2(+\infty)$
are the so-called fundamental solutions in the corresponding $n$-th SA
basis, i.e., the solutions for the initial conditions $b_1(0)=1$ and
$b_2(0)=0$ at $t=0$.
As we can see from the table, in the different SA bases, the transition
probability is expressed in terms of the fundamental solutions differently.
Moreover, depending on the SA basis, the desired probability for
nonadiabatic transitions is equal to the probability of no transitions
(for $n=0,2,4,\ldots$) or to the transition probability (for $n=1,3,5,\ldots$).

The last case in Table \ref{Table}, in which both the ``coupling'' and
the ``detuning'' are odd functions of time and which is given for the
sake of completeness, is not interesting in the context of the present
paper but it represents an interesting effect called symmetry forbidden
transitions, in which the system returns to its initial state in the end
of the interaction \cite{Vitanov95}.
This return is only determined by the symmetry of the Hamiltonian
and does not depend on its particular details.
Moreover, it remains valid in the general $N$-state case as well.

\vspace*{6mm}
\begin{center}
\begin{tabular}{|l|l|l|l|} \hline
Symmetry & 
${\bf U}_n(+\infty,-\infty)$ & 
Cases &
$P$ \\
\hline \hline
$\ba{l} \Omega_n(-t)=\Omega_n(t)\\ \Delta_n(-t)=\Delta_n(t)\ea$ &
$\ba{l} U_{11}=b_1^2+b_2^{*2} \\
	U_{12} = 2i{\rm Im} (b_1b_2) \ea$ &
odd $n$ &
$\left| U_{12} \right|^2$ \\
\hline
$\ba{l}\Omega_n(-t)=\Omega_n(t)\\ \Delta_n(-t)=-\Delta_n(t)\ea$ &
$\ba{l} U_{11}=\left| b_1 \right|^2 - \left| b_2 \right|^2 \\
	U_{12} = -2 b_1 b_2^* \ea$ &
$n=0$ &
$\left| U_{11} \right|^2$ \\
\hline
$\ba{l}\Omega_n(-t)=-\Omega_n(t)\\ \Delta_n(-t)=\Delta_n(t)\ea$ &
$\ba{l} U_{11}=b_1^2-b_2^{*2} \\
	U_{12} = -2{\rm Re} (b_1b_2) \ea$ &
even $n$ &
$\left| U_{11} \right|^2$ \\
\hline
$\ba{l}\Omega_n(-t)=-\Omega_n(t)\\ \Delta_n(-t)=-\Delta_n(t)\ea$ &
$\ba{l} U_{11}=1 \\
	U_{12} = 0 \ea$ &
 &
 \\
\hline
\end{tabular}   
\end{center}

\begin{table}[h]
\caption{
The relations between the fundamental solutions $b_1(+\infty)$ and
$b_2(+\infty)$ in the $n$-th superadiabatic basis
and the probability $P$ for nonadiabatic transitions.
In the elements of the evolution matrix ${\bf U}_n(+\infty,-\infty)$ in the
$n$-th SA basis as well as in the fundamental solutions we have omitted
for simplicity the arguments $\pm\infty$ and the label $n$.
}
\label{Table}
\end{table}

\end{appendix}



\end{document}